\documentclass[doublecol,a4paper]{epl2} 

\title{Phase transition in a super superspin glass}
\shorttitle{Phase transition in a super superspin glass}

\author{R. Mathieu\inst{*,1}, J. A. De Toro\inst{2}, D. Salazar\inst{3}, S. S. Lee\inst{4}, J. L. Cheong\inst{4}, and P. Nordblad\inst{1}}

\shortauthor{R. Mathieu \etal}

\institute{
\inst{*} Electronic address: roland.mathieu@angstrom.uu.se\\
\inst{1} Department of Engineering Sciences, Uppsala University, Box 534, SE-751 21 Uppsala, Sweden\\
\inst{2} Instituto Regional de Investigaci\'on Cient\'ifica Aplicada (IRICA) and Departamento de F\'isica Aplicada, Universidad de Castilla-La Mancha, 13071 Ciudad Real, Spain\\
\inst{3} Departamento de F\'isica Aplicada, Universidad de Castilla-La Mancha, 13071 Ciudad Real, Spain\\
\inst{4} Institute of Bioengineering and Nanotechnology, 31 Biopolis Way, The Nanos, Singapore 138669, Singapore

}


\abstract{We here confirm the occurrence of spin glass  phase transition and extract estimates of associated critical exponents of a highly monodisperse and densely compacted system of bare maghemite nanoparticles. This system has earlier been found to behave like an archetypal spin glass, with e.g. a sharp transition from paramagnetic to non-equilibrium behavior, suggesting that this system undergoes a spin-glass phase transition at a relatively high temperature, $T_g$ $\sim$ 140 K.}

\begin{document}

\maketitle

\textbf{Introduction.} - Spin glasses remain omnipresent in Condensed Matter Physics and there have been many recent experimental and theoretical studies trying to understand and explain the physical properties of spin glasses\cite{refs-sg-new} or the glassy phases exhibited by some functional materials\cite{refs-othersg}. It has also been found that by increasing the magnetic interaction between magnetic nanoparticles, i.e. between superspins comprising hundreds or thousands of atomic spins, the usual superparamagnetic response of the system is slowed down into spin glass like dynamics\cite{refs-nano}. Interestingly, in spite of displaying spin-glass features such as ageing, memory and rejuvenation\cite{refs-sg-new}, those superspin glasses have qualitatively different dynamical properties. The temperature onset of non-equilibrium dynamics is not as sharp as in atomic systems, and the dynamics itself seems less sensitive to temperature\cite{refs-nano,petra}.\\
\indent An extremely monodisperse system of bare maghemite $\gamma$-Fe$_2$O$_3$ nanoparticles (average diameter $\sim$ 8 nm) was compacted into a dense disc with a filling factor of 67 \% and structurally characterised in detail \cite{JADT}. The magnetic response of the disc was also investigated, and it was suggested from dynamical scaling analysis that the system undergoes a spin-glass phase transition near $T_g$ $\sim$ 140 K. In the present letter, we investigate the effect of an applied magnetic field on the dynamic and static response of the material. A static scaling analysis of the non-linear susceptibility data confirms the spin glass phase transition and yields estimates of the critical exponents associated with the transition.\\

\textbf{Experimental.} -  The volume ac-susceptibility $\chi$ of the disc was recorded as a function of  temperature $T$ or dc magnetic field $H$ on a Superconducting Quantum Interference Device (SQUID) magnetometer from Quantum Design Inc.  The in-phase $\chi'(T,f)$ and out-of-phase $\chi''(T,f)$ components of the susceptibility were recorded using different amplitudes of the ac-excitation field $h$, while the frequency $f$ was fixed to 170 Hz.\\

\begin{figure}[h]
\begin{center}
\includegraphics[width=0.46\textwidth]{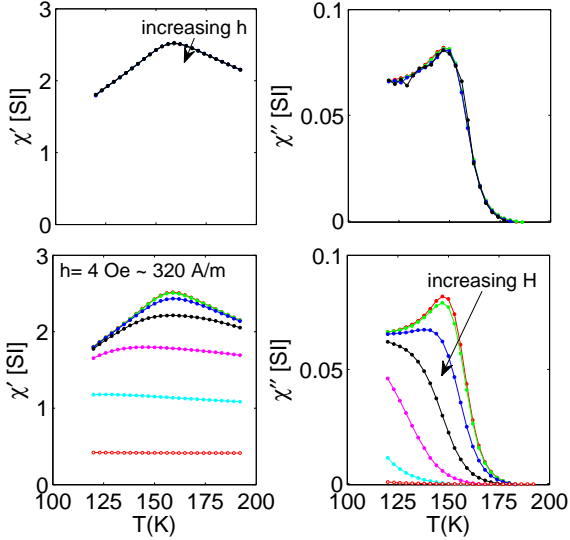}
\caption{Temperature dependence of the (left panels) in-phase component $\chi'$ and (right panels) out-of-phase component $\chi''$ of the ac-susceptibility recorded (upper panels) for different amplitudes of the ac excitation $h$ = 0.125, 0.4, 1.25 and 4 Oe ($\sim$ 10 to 80 A/m); lower panels: for a fixed $h$, and different superimposed dc bias fields $H$ = 0, 3, 10, 30, 100, 300, and 1000 Oe (0 to $\sim$ 80 kA/m). All data recorded on reheating with $f$ = 170 Hz.}
\label{fig1}
\end{center}
\end{figure}

\textbf{Results and discussion.} - The dependence of the ac-susceptibility on the amplitude of the ac-excitation $h$ and a superimposed dc magnetic field $H$ near the proposed spin-glass phase transition temperature is shown in Fig.~\ref{fig1}. As seen in the top panel of the figure, a linear magnetic response of the system is observed up to ac-excitations of 4 Oe ($\sim$ 320 A/m)\cite{orm}. Interestingly, this response is very sensitive to superimposed dc-fields, as illustrated in the lower panels of Fig.~\ref{fig1}. A small bias dc-field $H$ = 3 Oe ($\sim$ 240 A/m) is enough to affect the ac-susceptibility.  In $H$ = 1000 Oe ($\sim$ 80 kA/m), no dissipation appears in the investigated temperature interval. A similar sensitivity to superposed dc-fields has been observed in other superspin glasses\cite{petra}, while the application of larger magnetic fields tends to be necessary to influence the ac-susceptibility of (atomic) spin glasses\cite{petra-noTg}. 

\begin{figure}[h]
\begin{center}
\includegraphics[width=0.46\textwidth]{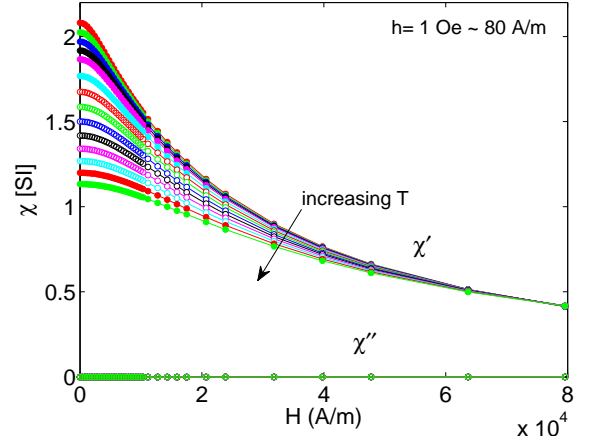}
\caption{In-phase component $\chi'$ and out-of-phase component $\chi''$ of the ac-susceptibility as a function of the dc bias field $H$ for different temperatures, from $T$ = 200 to 220 K in steps of 5 K, and 220 to 310 K in steps of 10 K. All data recorded on reducing the field from its maximum  value (1000 Oe $\sim$ 80 kA/m); $f$ = 170 Hz.}
\label{fig11}
\end{center}
\end{figure}

In order to extract the non-linear susceptibility of the material and investigate its divergence\cite{ising-stat,Levy}, the ac-susceptibility was measured as a function of dc bias field at temperatures  well above the freezing temperature, below which the system is out of equilibrium. The freezing temperature $T_f$($f$ = 170 Hz) $\sim$ 165 K is derived from the data presented in the upper panels of Fig.~\ref{fig1}  in the same way as  in the dynamical scaling analysis reported in ref. \cite{JADT}. Figure~\ref{fig11} shows the field dependence of both components of the susceptibilty for various temperatures, ranging from about 1.2 to 1.9 $T_f$. As expected and required there is no finite out-of phase component of the ac-susceptibility in this temperature region and no dynamical effects contribute to the measured response. 

\begin{figure}[h]
\begin{center}
\includegraphics[width=0.46\textwidth]{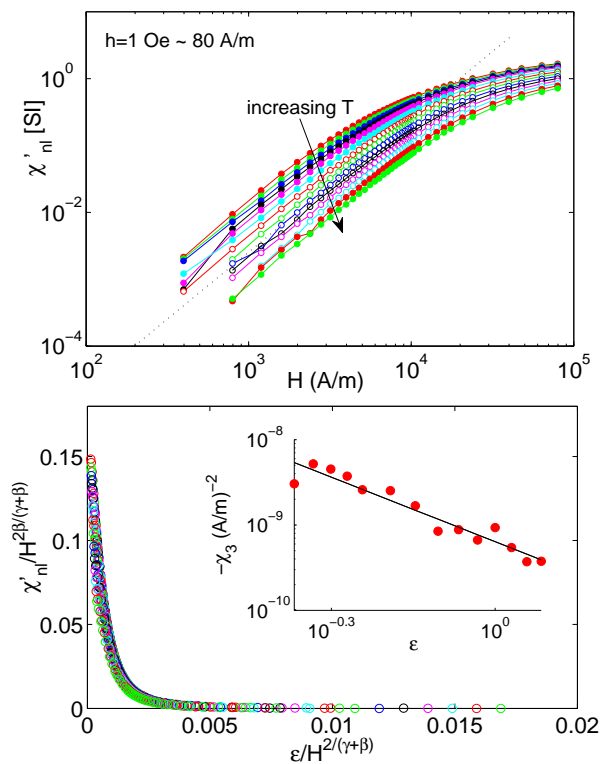}
\caption{Upper panel: non-linear susceptibility $\chi'_{nl}$ as a function of the dc bias field $H$ for different temperatures, extracted from the data in Fig.~\ref{fig11}. The dotted line indicates a $\chi'_{nl}$ $\propto$ $H^2$ behaviour (slope 2 in log-log plot) as a guide for the eye. Lower panel: The inset shows the scaling of $\chi_3$, the lowest order term of $\chi'_{nl}$, with the reduced temperature $\epsilon=(T-T_g)/T_g$. $T_g$ = 140 K obtained in \cite{JADT} is used. The main frame shows the scaling of the whole non linear susceptibility curves as $\chi'_{nl}/H^{2\beta/(\gamma + \beta)}$ versus $\epsilon/H ^{2/(\gamma + \beta)}$.}
\label{fig2}
\end{center}
\end{figure}

The magnetisation $M$ in a spin glass can be expressed as $M=\chi_1H+\chi_{nl}$. $\chi_1$ is the linear susceptibility and $\chi_{nl}$ the non-linear susceptibility, which includes the higher order terms:  $\chi_{nl}= \chi_3H^3+\chi_5H^5+...$. In a spin glass all the non-linear terms ($\chi_i$, $i$ $>$ 1) diverge at $T_g$\cite{susuki-barbara}. Hence the study of the divergence of the non-linear susceptibility can prove (or disprove) the occurrence of a spin glass phase transition. Rather than measuring those higher harmonics directly, it is convenient to extract the non-linear susceptibility data $\chi_{nl}=\chi_1-M/H$ from dc $M$ versus $H$  \cite{gingras-LHYF4} or ac-$\chi$ versus $H$ \cite{ising-stat,roland-xy} measurements. We can thus plot the data shown in Fig.~\ref{fig11} as $\chi'_{nl}=\chi(H=0)-\chi$ versus $H$ (in this temperature region, $\chi'$ = $\chi$).  This is shown in the upper panel of Fig.~\ref{fig2}. From this plot, one can extract the lowest order term of the non linear susceptibility $\chi_3$ for the various temperatures considered. It can be shown that for the smallest bias fields (still much larger than the amplitude of the ac-excitation $h$), $\chi'_{nl} \sim -3\chi_3H^2$\cite{Levy}. The obtained $\chi_3$($T$) data can be scaled with the reduced temperature $\epsilon = (T-T_g)/T_g$ as $\chi_3 \propto \epsilon^{-\gamma}$\cite{susuki-barbara}, as in the inset of Fig.~\ref{fig2}, evidencing the divergence of $\chi_3$ at $T_g$ = 140 K. The scaling yields for the spin glass susceptibility exponent $\gamma$ (thus defined as $\chi$ $\propto$ $\epsilon^{-\gamma}$) a value of 2.5 $\pm$ 0.5. 

\begin{table}[h]
\caption{Selected critical exponents for various three-dimensional spin glasses:  (atomic) Heisenberg-like Ag(Mn)\cite{Levy}, XY-like Eu$_{0.5}$Sr$_{1.5}$MnO$_4$\cite{roland-xy}, and Ising-like Fe$_{0.5}$Mn$_{0.5}$TiO$_3$\cite{ising-stat} spin glasses, interacting Fe-C nanoparticle system behaving as a superspin glass (SSG)\cite{ssg-stat}, the present super superspin glass (SSSG) system.}
\label{table1}
\begin{center}
\begin{tabular}{rcc}
& $\gamma$ & $\beta$ \\
\hline
SSG & 4 & 1.2 \\
\hline
Ising SG & 4 & 0.54 \\
XY SG & 3 & 0.5 \\
Heisenberg SG & 2.3 & 0.9 \\
\hline
\textbf{SSSG} & 2.5 & 0.2

\end{tabular}
\end{center}
\end{table}

There are several critical exponents describing the second-order phase transition such as the magnetisation exponent $\beta$ ($M$ $\propto$ $-\epsilon^\beta$) or heat capacity exponent ($C$ $\propto$ $\epsilon^{-\alpha}$).  The complete non-linear susceptibility curves can be scaled in the critical region as $\chi_{nl} = H^{2\beta/(\gamma + \beta)} G[\epsilon/H ^{2/(\gamma + \beta)}]$ where $G$ is a functional form\cite{susuki-barbara}. A relatively good collapse of the $\chi_{nl}(T,H)$ data may be obtained using $T_g$ = 140 K and $\gamma$ = 2.5, yielding $\beta$ = 0.2 $\pm$ 0.1. That scaling, like the power law dependence of $\chi_3$ on the reduced temperature, evidences the divergence of the spin glass susceptibility and the phase transition, in spite of the relatively large uncertainties on $\gamma$ and $\beta$. Typical values for those exponents are listed in Table~\ref{table1}. 

There is a relatively large variation of the various values reported in the literature, and in the table we have selected values which were obtained from static scaling analysis, that have been performed under correct conditions as regard temperature and magnetic field (critical regime)\cite{mattsson,ssg-stat}.  It is seen that the values of the exponents obtained  in the present case are similar to those reported on atomic spin glasses. Interestingly, it has also been found that the dynamical properties of the system are more similar to those of atomic spin glasses than those of superspin glasses (weakly accumulative ageing / significant rejuvenation)\cite{JADT}.\\

\textbf{Conclusion.} - We have  confirmed the existence of a spin glass phase transition in our densely compacted maghemite nanoparticle system. Hence remarkably, this system of superspins behaves like an archetypal spin glass, composed of atomic spins. This suggests that this ``super superspin glass'' could be used as a model system to test spin glass theories. Since the glassiness sets in at relatively large temperatures ($\sim$ 140 K) and the system is quite sensitive to magnetic field, in-field dynamical scaling experiments\cite{petra-noTg}, or detailed studies of the memory/rejuvenation phenomena\cite{memory} could readily be performed to  increase our understanding of the spin glass phase.\\

\acknowledgments
We thank the Swedish Research Council (VR), the G\"oran Gustafsson Foundation, and the Spanish CICYT [MAT2011-26207] for financial support.

\end{document}